\documentclass[a4paper]{jpconf}
\usepackage{graphicx}

\newcommand{\binds}{{\asymp}}
\newcommand{\dt}{{\rmd t}}

\newcommand{\node}{\circle{25}}
\newcommand{\blacknode}{\circle*{25}}
\newcommand{\smallnode}{\circle{12}}

\newcommand{\bra}{\langle}
\newcommand{\ket}{\rangle}

\newcommand{\order}{{\mathcal O}}

\newcommand{\bnull}{{\mbox{\boldmath $0$}}}
\newcommand{\be}{\begin{equation}}
\newcommand{\ee}{\end{equation}}
\newcommand{\bd}{\begin{displaymath}}
\newcommand{\ed}{\end{displaymath}}

\newcommand{\bigroom}{\rule[-0.4cm]{0cm}{1.0cm}}

\newcommand{\F}{{\mathcal F}}

\newcommand{\R}{{\rm I\!R}}

\newcommand{\bc}{\ensuremath{\mathbf{c}}}

\newcommand{\bk}{\ensuremath{\mathbf{k}}}

\newcommand{\bx}{\ensuremath{\mathbf{x}}}
\newcommand{\by}{\ensuremath{\mathbf{y}}}

\newcommand{\bpsi}{{\mbox{\boldmath $\psi$}}}

\newcommand{\bomega}{{\mbox{\boldmath $\omega$}}}

\begin{document}
\title{Generating functional analysis of complex formation and dissociation in large protein interaction networks}

\author{ACC Coolen and S Rabello}

\address{Department of Mathematics, King's College London, The Strand, London WC2R 2LS, U.K.\\
and Randall Division of Cell and Molecular Biophysics,
King's College London, New Hunt's House, London SE1 1UL, UK}

\ead{ton.coolen@kcl.ac.uk}

\begin{abstract}
We analyze large systems of interacting proteins, using techniques from the non-equilibrium statistical mechanics of disordered many-particle systems. Apart from protein production and removal, the most relevant microscopic processes in the proteome are complex formation and dissociation, and the microscopic degrees of freedom are the evolving concentrations of unbound proteins (in multiple post-translational states) and of protein complexes. Here we only include dimer-complexes, for mathematical simplicity, and we draw the network that describes which proteins are reaction partners from an ensemble of random graphs with an arbitrary degree distribution. We show how generating functional analysis methods
can be used successfully to derive closed equations for dynamical order parameters, representing an exact macroscopic description of the complex formation and dissociation dynamics in the infinite system limit. We end this paper with a discussion of the possible routes towards solving the nontrivial order parameter equations, either exactly (in specific limits) or approximately.
\end{abstract}

\section{Introduction}

 It is safe to say that biomedicine is undergoing a profound and irreversible change, brought about by what some coined a `data tsunami'. A spectacular increase of both the quality and quantity of data on cellular signalling, molecular structure and gene expressions, including {\em in vivo} data, have prompted biomedicine to transform rapidly into a much more quantitative as opposed to descriptive science, and
 theory now lags behind experiment to an almost embarrassing extent. The biological systems thought to be responsible for generating these data tend to be complex and nonlinear, involving many variables and nested processes acting on widely separated timescales. Also, biology lacks many of the simplifying features of many large physical systems, such as detailed balance, the identical nature of the degrees of freedom, or translation invariance. In biological systems, the interacting objects and their mutual forces are as a rule non-identical; in biology this would be called `inhomogeneity', where in physics one would speak of `disorder'.

One particular biological system of great relevance to medicine is the proteome, the collection of about $2.10^4$ cellular proteins and their molecular complexes, which constitute the main work force of the cell, responsible for executing most of the tasks that allow it to function, and for inter- and intra-cellular communication. The vast increase of proteomic data allow us to refine our understanding of its pathways and reactions, but we are running into new barriers: `The most significant challenges that the mechanistic modelling might be facing are the lack of quantitative kinetic data and the combinatorial increase in the number of emerging distinct species and states of the protein network being simulated' \cite{Kholodenko}.
Our aim in this paper is to contribute to our quantitative understanding of the proteome, by exploring the potential of quantitative macroscopic regularities emerging in very large as opposed to small proteomic systems. To do so we construct dynamical single-compartment models of large systems of interacting proteins, and analyze these in the limit of an infinite number of reaction partners, adapting techniques from the non-equilibrium statistical mechanics of disordered physical many-particle systems.

\section{Model definitions}

We imagine a cell in which  $N$ different proteins can be expressed, labeled by Roman indices $i\in\{1,\ldots, N\}$.
Each protein can be in at most $q$ post-translational states, labeled by Greek indices $\alpha\in\{1,\ldots,q\}$.
 The concentration in the cell of $(i,\alpha)$, the $\alpha$-th post-translational state of protein $i$, will be written as $x^\alpha_i$.
There are in principle $\frac{1}{2}N(N+1)$ possible complexes that these proteins could form by pairwise binding, $\frac{1}{2}N(N-1)$ hetero-dimers and $N$ homo-dimers. We denote the complex where $i$ binds to $j$ as $(i\binds j)$, and the concentration in the cell of the
complex $(i\binds j)$ as $x_{ij}$.
For mathematical convenience we can always take $x_{ij}=x_{ji}$.

\subsection{Elementary processes and dynamical equations}

  We aim to construct and analyze single-compartment type dynamical equations for the concentrations $\{x^\alpha_i\}$ of unbound proteins and $\{x_{ij}\}$ of their binary complexes, that capture the following elementary proteomic  processes:\vspace*{1mm}

  \begin{center}{
  \begin{tabbing}
  zzzzzzzzzzzzzzz\=zz\=zzzzzzzzzzzzzzzzzzzzzzzzzzzzzzzzzz\=zzzzzzzzzzzzzzzzzzzzzzzzzz\=\kill
  \>\> {\em elementary process:}                \> {\em symbolic notation:}   \> {\em process rate:}\\[1.5mm]
\>  \> binary complex formation:               \> $(i,\alpha)+(j,\beta)\to (i\binds j)$   \>   $k_{ij}^{\alpha\beta +}x^\alpha_i x^\beta_j$ \\[0.5mm]
\>  \> binary complex dissociation:            \> $(i\binds j)\to (i,\alpha)+(j,\beta)$   \>   $k_{ij}^{\alpha\beta -}x_{ij}$  \\[0.5mm]
\>  \> protein degradation/removal:     \> $(i,\alpha)\to \emptyset$  \>   $\gamma^\alpha_i x^\alpha_i$      \\[1mm]
\>  \> protein synthesis:              \> $\emptyset\to(i,\alpha)$   \>   $\theta^\alpha_i$          \\[-2mm]
\end{tabbing}
}\end{center}

\noindent
All rate parameters $\{k^{\alpha\beta\pm}_{ij}\!,\gamma^\alpha_i,\theta^\alpha_i\}$ are by definition non-negative. Of these parameters only $\{\gamma^\alpha_i,\theta^\alpha_i\}$ are allowed to depend on time, in order to incorporate possible changes in membrane properties and gene expression levels; the complex formation/dissociation parameters $\{k^{\alpha\beta\pm}_{ij}\}$, in contrast, depend only on the
structural characteristics of the reaction partners, via the indices $(i,\alpha)$ and $(j,\beta)$.
Consistency demands that $k^{\alpha\beta\pm}_{ij}=k_{ji}^{\beta\alpha\pm}$.
Not all pairs $(i,j)$ can in practice form a stable complex $(i\binds j)$, so we need (fixed) variables $c_{ij}\in\{0,1\}$ to define which
are potential binding partners: if $(i\binds j)$ is possible we put $c_{ij}\!=\!c_{ji}\!=\!1$, otherwise
$c_{ij}\!=\!c_{ji}\!=\!0$. The non-directed graph $\bc=\{c_{ij}\}$ thus defines the cell's protein interaction network.

Combining the four elementary processes above with the structural information in $\bc$ regarding
binding partners, we are then led to the following set of Michaelis-Menten reaction equations:
\begin{eqnarray}
\frac{\rmd}{\rmd t}x^\alpha_i&=& \sum_{j}c_{ij}\sum_\beta[k_{ij}^{\alpha\beta -}x_{ij}\!-\!k_{ij}^{\alpha\beta +}x^\alpha_i x^\beta_j]+\theta^\alpha_i-\gamma^\alpha_i x^\alpha_i
\label{eq:ddt1}
\\
\frac{\rmd}{\rmd t}x_{ij}&=& c_{ij}\sum_{\alpha\beta}[k_{ij}^{\alpha\beta +}x^\alpha_i x^\beta_j-k_{ij}^{\alpha\beta -}x_{ij}]
\label{eq:ddt2}
\end{eqnarray}
If $c_{ij}=0$ one will indeed have $x_{ij}=0$ at all times, as one should.
Equations (\ref{eq:ddt1},\ref{eq:ddt2}) satisfy the relevant mass conservation constraints:
the total amount of any protein $i$ (whether bound in dimers or in unbound form) changes only due  to production/degradation imbalance, viz.
\begin{eqnarray}
\frac{\rmd}{\rmd t}\Big[\sum_\alpha x_i^\alpha+\sum_j x_{ij}\Big]&=&
\sum_\alpha(\theta^\alpha_i-\gamma^\alpha_i x^\alpha_i)
\label{eq:conservation}
\end{eqnarray}
The presence of concentrations $x_{ij}$ of complexes, with their two protein indices, could complicate a direct statistical mechanical analysis
of equations (\ref{eq:ddt1},\ref{eq:ddt2}). However, the $x_{ij}$ are seen to obey linear equations, which we can simply solve:
\begin{eqnarray}
x_{ij}(t)&=& c_{ij}\sum_{\rho\lambda}k_{ij}^{\rho\lambda +}\int_{-\infty}^t\!\rmd s~\rme^{-k_{ij}^{-}(t-s)}x^\rho_i(s)x^\lambda_j(s)
\label{eq:eliminate}
\end{eqnarray}
with the short-hand $k_{ij}^{-}=\sum_{\alpha\beta}k_{ij}^{\alpha\beta -}\!$.
If we substitute (\ref{eq:eliminate}) into equations (\ref{eq:ddt1})  we obtain
\begin{eqnarray}
\frac{\rmd}{\rmd t}x^\alpha_i(t)
&=& \sum_{j}c_{ij}\int\!\rmd s~\sum_{\rho\lambda} W_{\alpha;\rho\lambda}(t\!-\!s|\bk_{ij})x^\rho_i(s)x^\lambda_j(s)
+\theta^\alpha_i(t)-\gamma^\alpha_i(t) x^\alpha_i(t)
\label{eq:microdynamics}
\end{eqnarray}
with a partially retarded effective free protein interaction kernel
\begin{eqnarray}
W_{\alpha;\rho\lambda}(\tau|\bk)=k^{\rho\lambda +}\Big[
\sum_\beta k^{\alpha\beta -}\theta[\tau]\rme^{-k^{-}\tau}-\delta_{\alpha\rho}\delta(\tau)\Big]
\end{eqnarray}
which obeys $\sum_\alpha\int\!\rmd\tau~W_{\alpha;\rho\lambda}(\tau|\bk)=0$, for all $(\rho,\lambda)$.

 The above definitions
 involve several simplifications, the main ones being that the only complexes are dimers, and that we disregard all spatial variations of protein and complex concentrations.
However, we would argue that our definitions incorporate the main events in the proteome, while still allowing (as we will see)  for an exact generating functional analysis
in the limit $N\to\infty$ (under suitable conditions on the statistical properties of the model parameters).

\subsection{Statistics of microscopic parameters - interaction network and reaction rates}

It is not possible at present to solve equations such as (\ref{eq:microdynamics}) analytically for arbitrary choices of the
microscopic model parameters $\{c_{ij},k_{ij}^{\alpha\beta\pm},\theta_i^\alpha,\gamma_i^\alpha\}$.
Neither can we benefit from built-in parameter regularities or symmetries (in contrast to e.g.
physical systems on lattices). This leaves the route of disordered systems theory, where one exploits the property that large many-particle systems often exhibit universal behaviour of macroscopic observables,
whose values will in the infinite system size limit
depend only on the {\em statistics} of the microscopic parameters rather than their precise realization.
It follows that these values can then be calculated by performing suitable averages over all microscopic parameter realizations with the correct statistics.

Thus we take the parameters $\{c_{ij},k_{ij}^{\alpha\beta\pm}\}$ to be generated randomly from some appropriate distribution,
which should ideally incorporate as much of our available biological information as possible.
We will in this paper draw the protein interaction network $\bc$ at random from an ensemble of random graphs with prescribed degrees $(k_1,\ldots,k_N)$,  that are in turn drawn randomly from an as yet unspecified degree distribution $p(k)$:
 \begin{eqnarray}
  P(\bc)&=&\frac{1}{Z}\prod_i\delta_{k_i,\sum_{j\neq i} c_{ij}}.
  \prod_i\Big[c_0\delta_{c_{ii},1}+(1\!-\!c_0)\delta_{c_{ii},0}\Big]
  \label{eq:ensemble}
  \end{eqnarray}
   The self-connections $c_{ii}$ are included to allow for homo-dimers (with $c_0\in[0,1]$ representing the fraction of proteins that can form homo-dimers), $\bra k\ket=\sum_k p(k)k>0$
   gives the average number of hetero-dimer binding partners per protein, and $Z$ is a normalization constant. In (\ref{eq:ensemble}) all graphs that exhibit the enforced degree sequence $(k_1,\ldots,k_N)$ are generated equally likely.

We similarly draw for each index pair $(i,j)$ with $i\leq j$ the complex formation/dissociation rates  $\bk_{ij}=\{k_{ij}^{\alpha\beta \pm}\}$
randomly and independently from a joint distribution $P(\bk)$ (we can use the same symbol $P$ here as in (\ref{eq:ensemble}), as the arguments will prevent ambiguity). Averages over $P(\bk)$ will be written as
$\int\!\rmd\bk~P(\bk)f(\bk)=\bra f(\bk)\ket_{\bk}$.
Although the rates for reactions involving one protein pair $(i,j)$ with $i\leq j$ are now statistically independent from those involving any other pair $(k,\ell)$ with $k\leq \ell$, we do not assume statistical independence at the level of post-translational states or off-rates versus on-rates, i.e.
we do {\em not} assume that $P(\bk)=\prod_{\alpha\beta}[P(k^{\alpha\beta +})P(k^{\alpha\beta -})]$. There are only two internal symmetries we have to insist on. First, when referring to homo-dimers we must have $k^{\alpha\beta\pm}=k^{\beta\alpha\pm}$, for reasons of consistency.
Second, in the case of hetero-dimers,  if we define $S\bk$ via $(S\bk)^{\alpha\beta\pm}=k^{\beta\alpha\pm}$ then we will require that $P(S\bk)=P(\bk)$ for all $\bk$, for technical reasons that will become clear later. This implies that the {\em probability} of assigning specific on/off rates to any reaction $(i,\alpha)+(j,\beta)\leftrightarrow i\binds j$ is the same as the probability to find these rates for the reaction  $(i,\beta)+(j,\alpha)\leftrightarrow i\binds j$, but not that the actual rates are themselves identical.
We note that, although the number $q$ of {\em potential} post-translational states labeled by $\alpha=1\ldots q$ is the same for all proteins, the {\em actual} number of post-translations states can be made to vary from one protein to another, by appropriate choices for the rate statistics $P(\bk)$.

The production and decay rates $(\theta_i^\alpha,\gamma_i^\alpha)$ need not be drawn randomly for our methods to apply
(as these have only a single protein index $i$), although one could do so for mathematical convenience. Here we will allow the rates $(\theta_i^\alpha,\gamma_i^\alpha)$ to be time-dependent, to incorporate the effects of proteome perturbations or receptor triggering.
If each protein is always assembled in one unique conformation, we must choose the production rates such that for each $i$ only one index $\alpha$ can have $\theta_i^\alpha\neq 0$ (representing the protein's native state).

\section{Generating functional analysis}

In the spirit of \cite{DeDominicis} we define a generating functional that allows us to calculate the time-dependent statistics of unbound protein concentrations; from these follow, via (\ref{eq:eliminate}), also those of all protein complexes. This functional $Z[\bpsi]$ is a straightforward generalization to time dependent stochastic variables of the conventional moment generating function of random variables.

\subsection{The disorder-averaged generating functional}

 We first discretize time according to $t_i=i\Delta$, with an elementary time step $\Delta$ that will be sent to zero, and use $\delta$-functions to enforce at each time the validity of  the equations (\ref{eq:microdynamics}):
\begin{eqnarray}
Z[\bpsi]&=& \int\!\Big[\prod_{i\alpha t} \rmd x^\alpha_i(t)~\rme^{\rmi\Delta \psi^\alpha_i(t)x^\alpha_i(t)}\Big]
\nonumber
\\
&&\hspace*{-13mm}\times \prod_{i\alpha t}\delta\Big[x^\alpha_i(t\!+\!\Delta)-x^\alpha_i(t)-\Delta\Big( \sum_{j}c_{ij}\int\!\rmd s\sum_{\rho\lambda} W_{\alpha;\rho\lambda}(t\!-\!s|\bk_{ij})x^\rho_i(s)x^\lambda_j(s)
+\theta^\alpha_i(t)\!-\!\gamma^\alpha_i(t) x^\alpha_i(t)\Big)\Big]
\nonumber
\\
&=&\int\!\prod_{i\alpha t}\Big[\frac{\rmd x^\alpha_i(t)\rmd \hat{x}^\alpha_i(t)}{2\pi}~\rme^{\rmi\Delta\psi^\alpha_i(t)x^\alpha_i(t)+\rmi\hat{x}^\alpha_i(t)
\big(x^\alpha_i(t+\Delta)-x^\alpha_i(t)-\Delta[\theta^\alpha_i(t)-\gamma^\alpha_i(t) x^\alpha_i(t)]\big)}\Big].\rme^{\Xi[\bc,\{\bk\}]}
\label{eq:Zfree}
\end{eqnarray}
in which only the following object depends on the network $\bc$ and the process rates $\{\bk\}$:
\begin{eqnarray}
\Xi[\bc,\{\bk\}]&=& -\rmi\Delta\sum_{ij} c_{ij} \sum_{\alpha\rho\lambda}\sum_t \hat{x}_i^\alpha(t) \int\!\rmd s~ W_{\alpha;\rho\lambda}(t\!-\!s|\bk_{ij})x^\rho_i(s)x^\lambda_j(s)
\nonumber
\\
&=& -\rmi \sum_i c_{ii}\Xi_{ii}(\bk_{ii})-\rmi \sum_{i<j}c_{ij}\Xi_{ij}(\bk_{ij})
\label{eq:Xi}
\end{eqnarray}
with
\begin{eqnarray}
\Xi_{ii}(\bk)
&=&\sum_{\alpha\rho\lambda}\int\!\rmd t\rmd\tau~W_{\alpha;\rho\lambda}(\tau|\bk)
 \hat{x}_i^\alpha(t) x^\rho_i(t\!-\!\tau)x^\lambda_i(t\!-\!\tau)
 \label{eq:Xi_ii}
\\
\Xi_{ij}(\bk)
&=&
\int\!\rmd t\rmd\tau \sum_{\rho\lambda} k^{\rho\lambda +} x^\rho_i(t\!-\!\tau)x^\lambda_j(t\!-\!\tau)
\nonumber
\\
&&\times
 \Big\{
\theta[\tau]\rme^{-k^{-}\tau}\sum_{\alpha\beta}k^{\alpha\beta -}[\hat{x}_i^\alpha(t) \!+\!\hat{x}_j^{\beta}(t)]
-\delta(\tau)[
\hat{x}_i^\rho(t)
\!+\! \hat{x}_j^\lambda(t)]\Big\}
\label{eq:Xi_ij}
\end{eqnarray}
To arrive at (\ref{eq:Xi},\ref{eq:Xi_ii},\ref{eq:Xi_ij}) we have used $k_{ij}^{\alpha\beta\pm}=k_{ji}^{\beta\alpha\pm}$.
Note that $Z[\bnull]=1$, by construction.

We next average the generating functional over the disorder, i.e. over the randomly generated interaction networks (\ref{eq:ensemble}) and reaction rates $\{\bk_{ij}\}$. The result will be written as $\overline{Z[\bpsi]}$. The idea behind definition (\ref{eq:Zfree}) is that calculating
$\overline{Z[\bpsi]}$ gives us access to
averages of observables that evolve according to the equations (\ref{eq:microdynamics})
without actually solving these equations, e.g.
\begin{eqnarray}
\overline{x_i^\alpha(t)}&=& -\lim_{\bpsi\to \bnull}\lim_{\Delta\to 0}\frac{\rmi}{\Delta}  \frac{\partial}{\partial\psi_i^\alpha(t)}\overline{Z[\bpsi]}
\\
\overline{x_i^\alpha(t) x_j^\beta(t^\prime)}&=& -\lim_{\bpsi\to \bnull}\lim_{\Delta\to 0}
\frac{1}{\Delta^2}  \frac{\partial^2}{\partial\psi_i^\alpha(t)\partial\psi_j^\beta(t^\prime)}\overline{Z[\bpsi]}
\end{eqnarray}
The disorder occurs only in $\Xi[\bc,\{\bk\}]$, so we have to calculate $\overline{\exp\Xi[\bc,\{\bk\}]}$.
For the present reaction rates, drawn independently for each $(i,j)$ with $i\leq j$, we may use the fact that in (\ref{eq:ensemble})
the homo-dimer
entries $c_{ii}$ are statistically independent of those representing hetero-dimers:
\begin{eqnarray}
\overline{\rme^{\Xi[\bc,\{\bk\}]}}&=&
\overline{\rme^{-\rmi \sum_i c_{ii}\Xi_{ii}(\bk_{ii})}}~.~
\overline{\rme^{-\rmi \sum_{i<j}\!c_{ij}\Xi_{ij}(\bk_{ij})}}
\end{eqnarray}
We tackle the two factors separately, using the fact that
the relevant matrices have entries $\Xi_{ij}(\bk_{ij})=\order(N^0)$. The first factor gives a simple expression which factorizes over the proteins:
\begin{eqnarray}
\overline{\rme^{-\rmi \sum_i c_{ii}\Xi_{ii}(\bk_{ii})}}&=& \prod_i\Big[1-c_0+ c_0\bra \rme^{-\rmi \Xi_{ii}(\bk)}\ket_{\bk}\Big]
\label{eq:Xi_ii_av}
\end{eqnarray}
The second factor contains the main complications. To average over the random interaction networks we first use the fact that a mathematically equivalent way to write (\ref{eq:ensemble}) is
\begin{eqnarray}
  P(\bc)&=&\frac{1}{\cal Z}\prod_i\delta_{k_i,\sum_{j\neq i} c_{ij}}.
  \prod_i\Big[c_0\delta_{c_{ii},1}+(1\!-\!c_0)\delta_{c_{ii},0}\Big].
  \prod_{i<j}\Big[\frac{\bra k\ket}{N}\delta_{c_{ij},1}+\Big(1\!-\!\frac{\bra k\ket}{N}\Big)\delta_{c_{ij},0}\Big]
  \label{eq:new_ensemble}
  \end{eqnarray}
  The reason is that the extra Poissonnian factor can be written in terms of the average degree $\bra k\ket=N^{-1}\sum_i k_i$ (and hence absorbed in the normalization constant ${\cal Z}$), via the identity
  \begin{eqnarray}
  \prod_{i<j}\Big[\frac{\bra k\ket}{N}\delta_{c_{ij},1}+\Big(1\!-\!\frac{\bra k\ket}{N}\Big)\delta_{c_{ij},0}\Big]
  &=&  \Big(1\!-\!\frac{\bra k\ket}{N}\Big)^{\frac{1}{2}N(N-1)} \rme^{\frac{1}{2}N\bra k\ket\log[\bra k\ket/(N-\bra k\ket)]}
  \end{eqnarray}
  (since $k_i=\sum_{j\neq i}c_{ij}$ for all $i$, by virtue of the constrained network degrees).
We use integral representations to implement the degree constraints, i.e.
\begin{eqnarray}
\prod_i\delta_{k_i,\sum_{j\neq i}c_{ij}}=\prod_i\int_{-\pi}^\pi\!\frac{\rmd \omega_i}{2\pi}~\rme^{\rmi\omega_i(k_i-\sum_{j\neq i}c_{ij})}
=\int_{-\pi}^\pi\!\prod_i\Big[\frac{\rmd \omega_i}{2\pi}\rme^{\rmi\omega_ik_i}\Big]\rme^{-\rmi \sum_{i<j}c_{ij}(\omega_i +\omega_j)}
\end{eqnarray}
In combination these ingredients allow us to write
\begin{eqnarray}
\!\overline{\rme^{-\rmi \sum_{i<j}c_{ij}\Xi_{ij}(\bk_{ij})}}\!&=&
\frac{1}{\cal Z}\int_{-\pi}^\pi\!\prod_i\Big[\frac{\rmd \omega_i}{2\pi}\rme^{\rmi\omega_ik_i}\Big]
\prod_{i<j}\Big\bra\sum_{c_{ij}}\Big[\frac{\bra k\ket}{N}\rme^{-\rmi[\omega_i +\omega_j+\Xi_{ij}(\bk)]}\delta_{c_{ij},1}\!+\Big(1\!-\!\frac{\bra k\ket}{N}\Big)\delta_{c_{ij},0}\Big]
\Big\ket_{\!\bk}
\nonumber
\\
&=&
\frac{1}{\cal Z}\int_{-\pi}^\pi\!\prod_i\Big[\frac{\rmd \omega_i}{2\pi}\rme^{\rmi\omega_ik_i}\Big]
\prod_{i<j}\Big[1+\frac{\bra k\ket}{N}\Big(\rme^{-\rmi(\omega_i +\omega_j)}\bra \rme^{-\rmi\Xi_{ij}(\bk)}\ket_{\bk}-1\Big)\Big]
\nonumber
\\
&=&
\frac{1}{\cal Z}\int_{-\pi}^\pi\!\prod_i\Big[\frac{\rmd \omega_i}{2\pi}\rme^{\rmi\omega_ik_i}\Big]
\rme^{\frac{\bra k\ket}{N}\sum_{i<j}[\rme^{-\rmi(\omega_i +\omega_j)}\bra \rme^{-\rmi\Xi_{ij}(\bk)}\ket_{\bk}-1]+\order(N^0)}
\end{eqnarray}
We may now use $\Xi_{ij}(\bk)=\Xi_{ji}(S\bk)$ (with the previously introduced post-translational state swap operator $S$ defined via $(S\bk)^{\alpha\beta\pm}=k^{\beta\alpha\pm}$) and the convention that $P(S\bk)=P(\bk)$, which allows us to write
$\bra \rme^{-\rmi\Xi_{ij}(\bk)}\ket_{\bk}=\bra \rme^{-\rmi\Xi_{ji}(S\bk)}\ket_{\bk}=\bra \rme^{-\rmi\Xi_{ji}(\bk)}\ket_{\bk}$, and  symmetrize the summation $\sum_{i<j}$:
\begin{eqnarray}
\overline{\rme^{-\rmi \sum_{i<j}c_{ij}\Xi_{ij}(\bk_{ij})}}&=&
\frac{e^{-\frac{1}{2}\bra k\ket N}}{\cal Z}\int_{-\pi}^\pi\!\prod_i\Big[\frac{\rmd \omega_i}{2\pi}\rme^{\rmi\omega_ik_i}\Big]
\rme^{\frac{\bra k\ket}{2N}\sum_{ij}\rme^{-\rmi(\omega_i +\omega_j)}\bra \rme^{-\rmi\Xi_{ij}(\bk)}\ket_{\bk}+\order(N^0)}
\label{eq:Xi_ij_av}
\end{eqnarray}
Obtaining full factorization over protein variables in $\overline{Z[\bpsi]}$, in leading order in $N$, requires finding a way to disentangle the different protein variables that appear in the quantity
\begin{eqnarray}
\frac{1}{N^2}\sum_{ij}\rme^{-\rmi(\omega_i +\omega_j)}\bra \rme^{-\rmi\Xi_{ij}(\bk)}\ket_{\bk}
&=&
\\[-1mm]
&&\hspace*{-55mm}
\frac{1}{N^2}\sum_{ij}\rme^{-\rmi(\omega_i +\omega_j)}\Big\bra \rme^{-\rmi
\int\!\rmd t\rmd\tau \sum_{\rho\lambda} k^{\rho\lambda +} x^\rho_i(t-\tau)x^\lambda_j(t-\tau)
 \big\{
\theta[\tau]\rme^{-k^{-}\tau}\sum_{\alpha\beta}k^{\alpha\beta -}[\hat{x}_i^\alpha(t) +\hat{x}_j^{\beta}(t)]
-\delta(\tau)[
\hat{x}_i^\rho(t)
+ \hat{x}_j^\lambda(t)]\big\}}\Big\ket_{\!\bk}
\nonumber
\end{eqnarray}

\subsection{Introduction of dynamical order parameters}

To achieve factorization we introduce an appropriate dynamical order parameter and isolate the joint distribution of unbound protein
concentration `paths' $\{x_i^\alpha\}=\{x^\alpha_i(t)\}$ (with $t\in\R$), for each of their post-translational variants,
and their conjugate concentration paths:
\begin{eqnarray}
P[\{x,\hat{x}\}|\{\bx,\hat{\bx}\},\bomega]&=& \frac{1}{N}\sum_i
\prod_\alpha\delta[\{x_\alpha\}-\{x^\alpha_i\}] \delta[\{\hat{x}_\alpha\}-\{\hat{x}^\alpha_i\}]\rme^{-\rmi\omega_i}
\label{eq:orderparameter}
\end{eqnarray}
 With this macroscopic object we can write, upon defining the short-hand $\{\rmd x\}=\prod_{t\alpha} \rmd x_\alpha(t)$ and sending $\Delta\to 0$ wherever feasible,
 \begin{eqnarray}
\frac{1}{N^2}\sum_{ij}\rme^{-\rmi(\omega_i +\omega_j)}\bra \rme^{-\rmi\Xi_{ij}(\bk)}\ket_{\bk}
&=&
\int\{\rmd x\rmd\hat{x}\rmd x^\prime \rmd \hat{x}^\prime\}P[\{x,\hat{x}\}|\{\bx,\hat{\bx}\},\bomega]P[\{x^\prime\!,\hat{x}^\prime\}|\{\bx,\hat{\bx}\},\bomega]
\nonumber
\\
&&\hspace*{-50mm}\times
\Big\bra \rme^{-\rmi
\int\!\rmd t\rmd\tau \sum_{\rho\lambda} k^{\rho\lambda +} x_\rho(t-\tau)x^\prime_\lambda(t-\tau)
 \big\{
\theta[\tau]\rme^{-k^{-}\tau}\sum_{\alpha\beta}k^{\alpha\beta -}[\hat{x}_\alpha(t) +\hat{x}^\prime_{\beta}(t)]
-\delta(\tau)[
\hat{x}_\rho(t)
+\hat{x}^\prime_\lambda(t)]\big\}}\Big\ket_{\!\bk}
\end{eqnarray}
To transform $\overline{Z[\bpsi]}$ into an expression that can be evaluated for $N\to\infty$ by steepest descent,
we have to relocate the kernels $P[\{x,\hat{x}\}|\ldots]$, via the insertion (for each joint path $\{x,\hat{x}\}$)
of
\begin{eqnarray}
1&=& \int\!\rmd P[\{x,\hat{x}\}]\delta\Big[P[\{x,\hat{x}\}]-P[\{x,\hat{x}\}|\{\bx,\hat{\bx}\},\bomega]\Big]\nonumber
\\
&=& \int\!\frac{\rmd P[\{x,\hat{x}\}]\rmd \hat{P}[\{x,\hat{x}\}]}{2\pi/N}~\rme^{\rmi N\hat{P}[\{x,\hat{x}\}]\big[P[\{x,\hat{x}\}]-P[\{x,\hat{x}\}|\{\bx,\hat{\bx}\},\bomega]\big]}
\end{eqnarray}
After appropriate re-scaling of the conjugate kernels $\hat{P}[\{x,\hat{x}\}]$ (involving factors $N$ and $\Delta$) in order to obtain well-defined functional integrals with the correct $N$-scaling,
we then arrive at
\begin{eqnarray}
\rme^{\frac{\bra k\ket}{2N}\sum_{ij}\rme^{-\rmi(\omega_i +\omega_j)}\bra \rme^{-\rmi\Xi_{ij}(\bk)}\ket_{\bk}}
&=&
\int\!\{\rmd P\rmd \hat{P}\}
~\rme^{\rmi N\int\{\rmd x\rmd \hat{x}\}\hat{P}[\{x,\hat{x}\}]P[\{x,\hat{x}\}]}
\prod_i
~\rme^{-\rmi\hat{P}[\{x_i,\hat{x}_i\}]\rme^{-\rmi\omega_i}}
\nonumber
\\
&&
\hspace*{-30mm}
\times
\exp\left\{
\frac{1}{2}\bra k\ket N\!\int\!\{\rmd x\rmd\hat{x}\rmd x^\prime\! \rmd \hat{x}^\prime\}P[\{x,\hat{x}\}]P[\{x^\prime\!,\hat{x}^\prime\}]
\right.
\label{eq:X_ij_factorized}
\\
&&
\hspace*{-30mm}
\left.\times
\Big\bra \rme^{-\rmi
\int\!\rmd t\rmd\tau\! \sum_{\rho\lambda}\! k^{\rho\lambda +} x_\rho(t-\tau)x^\prime_\lambda(t-\tau)
 \big\{
\theta[\tau]\rme^{-k^{-}\tau}\sum_{\alpha\beta}\! k^{\alpha\beta -}[\hat{x}_\alpha(t) +\hat{x}^\prime_{\beta}(t)]
-\delta(\tau)[
\hat{x}_\rho(t)
+\hat{x}^\prime_\lambda(t)]\big\}}\Big\ket_{\bk}\right\}
\nonumber
\end{eqnarray}
Only the last factor of the first line contains the protein variables $\{x_i,\hat{x}_i,\omega_i\}$, and in a factorized form.
This allows us to combine our intermediate results (\ref{eq:Zfree},\ref{eq:Xi},\ref{eq:Xi_ii},\ref{eq:Xi_ij},\ref{eq:Xi_ii_av},\ref{eq:Xi_ij_av},\ref{eq:X_ij_factorized}) into
\begin{eqnarray}
\overline{Z[\bpsi]}&=& \frac{1}{\cal Z}\!
\int\!\{\rmd P\rmd \hat{P}\} \exp\Big\{ \order(N^0)+
~\rmi N\!\int\!\{\rmd x\rmd \hat{x}\}\hat{P}[\{x,\hat{x}\}]P[\{x,\hat{x}\}]-\frac{1}{2}N\bra k\ket
\nonumber
\\
&&
 +
\frac{1}{2}N\bra k\ket \!\!\int\!\{\rmd x\rmd\hat{x}\rmd x^\prime\! \rmd \hat{x}^\prime\}P[\{x,\hat{x}\}]P[\{x^\prime\!,\hat{x}^\prime\}]
\nonumber\\
&&
\hspace*{5mm}
\times
\Big\bra \rme^{-\rmi\!
\int\!\rmd t\rmd\tau\! \sum_{\rho\lambda}\! k^{\rho\lambda +} x_\rho(t-\tau)x^\prime_\lambda(t-\tau)
 \big\{
\theta[\tau]\rme^{-k^{-}\tau}\sum_{\alpha\beta}\! k^{\alpha\beta -}[\hat{x}_\alpha(t) +\hat{x}^\prime_{\beta}(t)]
-\delta(\tau)[
\hat{x}_\rho(t)
+\hat{x}^\prime_\lambda(t)]\big\}}\Big\ket_{\bk}\Big\}
\nonumber
\\
&&\times
\prod_i\left\{
\int_{-\pi}^\pi\!\frac{\rmd \omega}{2\pi}\rme^{\rmi\omega k_i}
\int\!\{\rmd x\rmd \hat{x}\}~\rme^{\rmi\int\!\rmd t\sum_\alpha \psi^\alpha_i(t)x_\alpha(t)+\rmi\int\!\rmd t \sum_\alpha\hat{x}_\alpha(t)
\big[\frac{\rmd}{\rmd t}x_\alpha(t)-\theta^\alpha_i(t)+\gamma^\alpha_i(t) x_\alpha(t)\big]}
\nonumber\right.
\\
&&\left.
\hspace*{10mm}
\times \rme^{-\rmi\hat{P}[\{x,\hat{x}\}]\rme^{-\rmi\omega}}
 \Big[1\!-c_0+ c_0\bra \rme^{-\rmi \sum_{\alpha\rho\lambda}\int\!\rmd t\rmd\tau~W_{\alpha;\rho\lambda}(\tau|\bk)
 \hat{x}_\alpha(t) x_\rho(t-\tau)x_\lambda(t-\tau)}\ket_{\bk}\Big]
\bigroom \right\}
\label{eq:factorized}
\end{eqnarray}
Now we
can for $N\to\infty$ carry out the functional integrations over $\{P,\hat{P}\}$ by steepest
 descent.
Upon defining $\hat{P}[\{x,\hat{x}\}]=\rmi Q[\{x,\hat{x}\}]$ the outcome takes the form
\begin{eqnarray}
\lim_{N\to\infty}\frac{1}{N}\log\overline{Z[\psi]}&=&
{\rm extr}_{\{P,Q\}}\Big\{\Psi[\{P,Q\}]+\Phi[\{P\}]+\Omega[\{Q\}]+{\rm const}\Big\}
\label{eq:saddlepointform}
\end{eqnarray}
where the constant follows from the identity $\log\overline{Z[\bnull]}=0$, and
with
\begin{eqnarray}
\Psi[\{P,Q\}]&=&
-\int\!\{\rmd x \rmd\hat{x}\} P[\{x,\hat{x}\}]Q[\{x,\hat{x}\}]-\frac{1}{2}\bra k\ket
\\
\Phi[\{P\}]&=& \frac{1}{2}\bra k\ket \!\!\int\!\{\rmd x\rmd\hat{x}\rmd x^\prime\! \rmd \hat{x}^\prime\}P[\{x,\hat{x}\}]P[\{x^\prime\!,\hat{x}^\prime\}]\times
\\
&&
\big\bra \rme^{-\rmi\!
\int\!\rmd t\rmd\tau\! \sum_{\rho\lambda}\! k^{\rho\lambda +} x_\rho(t-\tau)x^\prime_\lambda(t-\tau)
 \big\{
\theta[\tau]\rme^{-k^{-}\tau}\sum_{\alpha\beta}\! k^{\alpha\beta -}[\hat{x}_\alpha(t) +\hat{x}^\prime_{\beta}(t)]
-\delta(\tau)[
\hat{x}_\rho(t)
+\hat{x}^\prime_\lambda(t)]\big\}}\big\ket_{\bk}
\nonumber
\\
\Omega[\{Q\}]&=&
\Big\bra\log
\int_{-\pi}^\pi\!\frac{\rmd \omega}{2\pi}\rme^{\rmi\omega k}
\int\!\{\rmd x\rmd \hat{x}\}~\rme^{\rmi\int\!\rmd t\sum_\alpha\big[ \psi_\alpha(t)x_\alpha(t)+\hat{x}_\alpha(t)
\big(\frac{\rmd}{\rmd t}x_\alpha(t)-\theta_\alpha(t)+\gamma_\alpha(t) x_\alpha(t)\big)\big]}
\nonumber
\\
\hspace*{-15mm}&&
\hspace*{5mm}
\times \rme^{Q[\{x,\hat{x}\}]\rme^{-\rmi\omega}}
 \Big\bra \rme^{-\rmi s\sum_{\alpha\rho\lambda}\int\!\rmd t\rmd\tau~W_{\alpha;\rho\lambda}(\tau|\bk)
 \hat{x}_\alpha(t) x_\rho(t-\tau)x_\lambda(t-\tau)}\Big\ket_{\!\bk,s}
\Big\ket_{\!\{\psi,\theta,\gamma,k\}}
\end{eqnarray}
\noindent
with the short-hands
$\bra f[\{\psi_\alpha,\theta_\alpha,\gamma_\alpha\},k]\ket_{\{\psi,\theta,\gamma,k\}}= \lim_{N\to\infty}N^{-1}
\sum_i f[\{\psi_i^\alpha,\theta^\alpha_i,\gamma^\alpha_i\},k_i]$, and with $\bra f(s)\ket_s=c_0f(1)+(1-c_0)f(0)$. The random variable $s$
controls whether ($s=1$) or not ($s=0$) a given protein species can form homo-dimers.
The functional saddle-point equations that define the extremum in (\ref{eq:saddlepointform}), and are fully closed by definition, are the following:
\begin{eqnarray}
&&
\frac{\delta}{\delta P[\{x,\hat{x}\}]}\Psi[\{P,Q\}]+\frac{\delta}{\delta P[\{x,\hat{x}\}]}\Phi[\{P\}]=0
\label{eq:SP1}
\\
&&
\frac{\delta}{\delta Q[\{x,\hat{x}\}]}\Psi[\{P,Q\}]+\frac{\delta}{\delta Q[\{x,\hat{x}\}]}\Omega[\{Q\}]=0
\label{eq:SP2}
\end{eqnarray}
From now on we analyze  equations (\ref{eq:SP1},\ref{eq:SP2}) only for $\bpsi=\bnull$, the generating fields $\bpsi$ will be used in due course
only to identify the physical meaning of our order parameters.

\subsection{Saddle point equations}

For $N\to\infty$ the dynamical order parameters $P[\{x,\hat{x}\}]$ and $Q[\{x,\hat{x}\}]$ take well-defined  values that depend only on the statistical properties of the microscopic system parameters, and which follow upon solving the coupled equations (\ref{eq:SP1},\ref{eq:SP2}).
Working out these fully exact formulae gives:
\begin{eqnarray}
Q[\{x,\hat{x}\}]&=&
\bra k\ket \!\!\int\!\{\rmd x^\prime\! \rmd \hat{x}^\prime\}P[\{x^\prime\!,\hat{x}^\prime\}]\times
 \label{eq:SP1worked}
\\
&&
\big\bra \rme^{-\rmi\!
\int\!\rmd t\rmd\tau\! \sum_{\rho\lambda}\! k^{\rho\lambda +} x_\rho(t-\tau)x^\prime_\lambda(t-\tau)
 \big\{
\theta[\tau]\rme^{-k^{-}\tau}\sum_{\alpha\beta}\! k^{\alpha\beta -}[\hat{x}_\alpha(t) +\hat{x}^\prime_{\beta}(t)]
-\delta(\tau)[
\hat{x}_\rho(t)
+\hat{x}^\prime_\lambda(t)]\big\}}\big\ket_{\bk}
\nonumber
\\
P[\{x,\hat{x}\}]&=&
\Big\bra
\frac{\int_{-\pi}^\pi\!\frac{\rmd \omega}{2\pi}\rme^{\rmi\omega (k-1)+Q[\{x,\hat{x}\}]\rme^{-\rmi\omega}}
M[\{x,\hat{x},\theta,\gamma\}]
}
{
\int\!\{\rmd x^\prime\rmd \hat{x}^\prime\}
\int_{-\pi}^\pi\!\frac{\rmd \omega}{2\pi}\rme^{\rmi\omega k+Q[\{x^\prime\!,\hat{x}^\prime\}]\rme^{-\rmi\omega}}
M[\{x^\prime\!,\hat{x}^\prime,\theta,\gamma\}]
}
\Big\ket_{\!\{\theta,\gamma,k\}}
 \label{eq:SP2worked}
\end{eqnarray}
in which
\begin{eqnarray}
M[\{x,\hat{x},\theta,\gamma\}]&=& \Big\bra
\rme^{\rmi\int\!\rmd t\sum_\alpha\hat{x}_\alpha(t)
\big[\frac{\rmd}{\rmd t}x_\alpha(t)-\theta_\alpha(t)+\gamma_\alpha(t) x_\alpha(t)
-s\sum_{\rho\lambda}\int\!\rmd\tau~W_{\alpha;\rho\lambda}(\tau|\bk)
  x_\rho(t-\tau)x_\lambda(t-\tau)\big]}\Big\ket_{\!\bk,s}
 \nonumber \\
  &&
  \label{eq:Mmeasure}
\end{eqnarray}\vspace*{-6mm}

\noindent
The identity
$(2\pi)^{-1}\int_{-\pi}^\pi\!\rmd\omega~\exp[\rmi\omega k+z\rme^{-\rmi\omega}]=
 z^k/k!$ allows us to simplify (\ref{eq:SP2worked}) to
\begin{eqnarray}
P[\{x,\hat{x}\}]&=&
\Big\bra
\frac{kQ^{k-1}[\{x,\hat{x}\}]
M[\{x,\hat{x},\theta,\gamma\}]
}
{
\int\{\rmd x^\prime\rmd \hat{x}^\prime\}
Q^k[\{x^\prime\!,\hat{x}^\prime\}]
M[\{x^\prime\!,\hat{x}^\prime,\theta,\gamma\}]
}
\Big\ket_{\!\{\theta,\gamma,k\}}
 \label{eq:SP2worked_more}
\end{eqnarray}
It seems appropriate at this stage to switch from $P[\{x,\hat{x}\}]$  to its partial Fourier transform, which will lead to equations involving only real-valued order parameters that can be interpreted (as will be shown) as conditional path probabilities in the sense suggested by our adopted notation:
\begin{eqnarray}
W[\{x\}|\{y\}]&=& \int\!\{\rmd\hat{x}\} P[\{x,\hat{x}\}]\rme^{-\rmi\int\!\rmd t\sum_\alpha y_\alpha(t)\hat{x}_\alpha(t)}
\label{eq:WP}
\end{eqnarray}
Adopting this new definition converts our closed equations
(\ref{eq:SP1worked},\ref{eq:SP2worked_more}) into
\begin{eqnarray}
W[\{x\}|\{y\}]&=&
\Big\bra
\frac{k\int\!\{\rmd\hat{x}\} Q^{k-1}[\{x,\hat{x}\}]
M[\{x,\hat{x},\theta\!+\!y,\gamma\}]
}
{
\int\{\rmd x^\prime\rmd \hat{x}^\prime\}
Q^k[\{x^\prime\!,\hat{x}^\prime\}]
M[\{x^\prime\!,\hat{x}^\prime,\theta,\gamma\}]
}
\Big\ket_{\!\{\theta,\gamma,k\}}
 \label{eq:SPinW1}
\\
Q[\{x,\hat{x}\}]&=& \bra k\ket \int\{\rmd x^\prime \rmd y\}W[\{x^\prime\}|\{y\}]
\Big\bra \rme^{-\rmi\int\!\dt \sum_\alpha \hat{x}_\alpha(t)\int\!\rmd\tau \sum_{\rho\lambda}
W_{\alpha;\rho\lambda}(\tau|\bk)x_\rho(t-\tau)x_\lambda^\prime(t-\tau)}
\nonumber
\\
&&\times
\prod_{\alpha t}\delta\Big[y_\alpha(t)-
\int\!\rmd\tau \sum_{\rho\lambda}
W_{\alpha;\rho\lambda}(\tau|S\bk)x^\prime_\rho(t\!-\!\tau)x_\lambda(t\!-\!\tau)
\Big]\Big\ket_{\!\bk}
\label{eq:SPinW2}
\end{eqnarray}
with the previously introduced short-hand $(S\bk)^{\alpha\beta\pm}=k^{\beta\alpha\pm}$, and where we have used the following identity
which follows trivially from definition (\ref{eq:Mmeasure}),
\begin{eqnarray}
M[\{x,\hat{x},\theta,\gamma\}]\rme^{-\rmi\int\!\rmd t\sum_\alpha y_\alpha(t)\hat{x}_\alpha(t)}=
M[\{x,\hat{x},\theta\!+\!y,\gamma\}]
\end{eqnarray}
To work out equations (\ref{eq:SPinW1},\ref{eq:SPinW2}) further we need to evaluate the following integrals, for $k>0$:
\begin{eqnarray}
&&
\hspace*{-10mm}
\int\!\{\rmd\hat{x}\} Q^k[\{x,\hat{x}\}]
M[\{x,\hat{x},\theta,\gamma\}]
\nonumber
\\
&=& C\bra k\ket^k\int\prod_{\ell\leq k} \Big[\{\rmd x_\ell \rmd y_\ell\}W[\{x_\ell\}|\{y_\ell\}]
\Big\bra \prod_{\ell\alpha t}\delta\Big[y_{\ell\alpha}(t)-
\int\!\rmd\tau \sum_{\rho\lambda}
W_{\alpha;\rho\lambda}(\tau|S\bk_\ell)x_{\ell\rho}(t\!-\!\tau)x_{\lambda}(t\!-\!\tau)
\Big]
\nonumber
\\
&&\hspace*{3mm}\times\lim_{\Delta\to 0}
\prod_{\alpha t}\delta\Big[x_\alpha(t\!+\!\Delta)-x_\alpha(t)
-\Delta[\theta_\alpha(t)-\gamma_\alpha(t) x_\alpha(t)]
\nonumber
\\
&&\hspace*{6mm}
-\Delta\sum_{\rho\lambda}\int\!\rmd\tau~x_\rho(t\!-\!\tau)\Big(s W_{\alpha;\rho\lambda}(\tau|\bk)
  x_\lambda(t\!-\!\tau)\!+\!\sum_\ell
W_{\alpha;\rho\lambda}(\tau|\bk_\ell)x_{\ell\lambda}(t\!-\!\tau)\Big)
\Big]
\Big\ket_{\!\bk,s;\bk_1\ldots \bk_k}
\label{eq:basic_obstacle}
\end{eqnarray}
Here $C$ is a constant, which will drop out of our equations.
Further simplification of equations (\ref{eq:SPinW1},\ref{eq:SPinW2}) requires
knowing the physical interpretation of
the order parameter
$W[\{x\}|\{y\}]$.

\subsection{Physical meaning of order parameters}

We can infer the meaning of $W[\{x\}|\{y\}]$, starting from
(\ref{eq:orderparameter},\ref{eq:WP}) and using the manipulations applied to (\ref{eq:Zfree}),  which effectively boils down to making
in (\ref{eq:factorized}) the substitution
 \begin{eqnarray}
 \rme^{\rmi\!\int\!\rmd t\sum_{i\alpha}\psi^\alpha_i(t)x^\alpha_i(t)}~\to~ \frac{1}{N}\sum_j\rme^{-\rmi\omega_j}\prod_\alpha\Big[\delta[\{x_\alpha\}\!-\!\{x^\alpha_j\}]~\rme^{-\rmi\!\int\!\dt~ y_\alpha(t)\hat{x}^\alpha_j(t)}
 \Big]
 \vspace*{-1mm}
 \end{eqnarray}
Carrying out this substitution shows that insertion of the factor $\exp[-\rmi\omega_j]$ into (\ref{eq:factorized}) is equivalent to replacing $k_j\to k_j-1$, and that
insertion of the factor $\exp[-\rmi\!\int\!\dt~ y_\alpha(t)\hat{x}^\alpha_j(t)]$ is equivalent to replacing
 $\theta_j^\alpha(t)\to \theta_j^\alpha(t)+y_\alpha(t)$. We may therefore conclude that
  \begin{eqnarray}
W[\{x\}|\{y\}]
&=&
\frac{1}{N}\sum_{j}
\overline{\big\bra\delta[\{x\}-\{x_j\}]\big\ket}\big|_{k_j\to k_j-1,~~\theta^\alpha_j(t)\rightarrow \theta^\alpha_j(t)+y_\alpha(t)~\forall\alpha}
\label{eq:meaning}
 \vspace*{-1mm}
\end{eqnarray}
where the brackets $\bra \ldots\ket$ denote evaluation of the argument for the microscopic process (\ref{eq:microdynamics}).
$W[\{x\}|\{y\}]$ is apparently a generalized response function. It gives the probability that
if we pick at random a protein species $j$, remove one randomly selected binding partner from it and increase instead its production rates $\{\theta^\alpha_j\}$ by (time-dependent) amounts $\{y_\alpha\}$, we will observe for the post-translational states of that protein the concentration evolution $\{x_\alpha\}$. Similarly,
 \begin{eqnarray}
D[\{x\}|\{y\}]
&=&
\frac{1}{N}\sum_j
\overline{\big\bra\delta[\{x\}-\{x_j\}]\big\ket}\big|_{\theta^\alpha_j(t)\rightarrow \theta^\alpha_j(t)+y_\alpha(t)~\forall\alpha}
\label{eq:meaning2}
 \vspace*{-1mm}
\end{eqnarray}
(which quantifies the response to perturbation of the actual original system, without the node removals).
Since the calculation of (\ref{eq:meaning2}) differs from that of (\ref{eq:meaning}) only in the removal  of the factors
$\exp[-\rmi \omega_i]$ from definition (\ref{eq:orderparameter}),  tracing the
differences between the two calculations is easy. One finds that it boils down to replacing $\exp[\rmi\omega(k-1)]$ by $\exp[\rmi\omega k]$ in (\ref{eq:SP2worked}), and hence replacing  $kQ^{k-1}[\{x,\hat{x}\}]$ by $Q^{k}[\{x,\hat{x}\}]$ in (\ref{eq:SPinW1}),
which leads us directly to
\begin{eqnarray}
D[\{x\}|\{y\}]&=& \sum_{k\geq 0} p(k)
\Big\bra
\frac{\int\!\{\rmd\hat{x}\} Q^{k}[\{x,\hat{x}\}]
M[\{x,\hat{x},\theta\!+\!y,\gamma\}]
}
{
\int\{\rmd x^\prime\rmd \hat{x}^\prime\}
Q^k[\{x^\prime\!,\hat{x}^\prime\}]
M[\{x^\prime\!,\hat{x}^\prime,\theta,\gamma\}]
}
\Big\ket_{\!\{\theta,\gamma\}}
 \label{eq:Dexpression}
  \vspace*{-1mm}
\end{eqnarray}
Both $W[\{x\}|\{y\}]$ and $D[\{x\}|\{y\}]$ are conditional measures in the paths $\{x\}$, so
they must obey $\int\{\rmd x\}~W[\{x\}|\{y\}]=\int\{\rmd x\}~D[\{x\}|\{y\}]=1$ for all $\{y\}$.
After solving $W[\{x\}|\{y\}]$ from (\ref{eq:SPinW1},\ref{eq:SPinW2}), one obtains $D[\{x\}|\{y\}]$ via (\ref{eq:Dexpression}), and extracts $D[\{x\}|\{0\}]$ as the measure for the concentrations of unbound proteins in the unperturbed system.
It follows from (\ref{eq:meaning}), together with the causality of our equations (\ref{eq:microdynamics}) and the fact that we have discretized our microscopic laws according to the It\^{o} convention $\rmd x(t)=x(t+\rmd t)-x(t)$,
 that each $x_\alpha(t)$ in $W[\{x\}|\{y\}]$ can depend only on those $y_\beta(t^\prime)$ that have $t^\prime<t$ (to be precise: that have $t^\prime\leq t-\Delta$ before the $\Delta\to 0$ limit).

\subsection{Implications of causality}

Returning to (\ref{eq:basic_obstacle}), we see that causality allows us to integrate (\ref{eq:basic_obstacle}) over $\{x\}$. We first discretize time and integrate over the $x_\alpha(t)$ with $t=t_{\rm max}$, followed by integration over the $x_{\ell\alpha}^\prime(t_{\rm max})$ and the $y_{\ell\alpha}(t_{\rm max})$ (in that order). The net result is the same expression we started out with, but with $t_{\rm max}$ replaced by $t_{\rm max}-\Delta$. Repetition of the argument then leads to the simple result
\begin{eqnarray}
\int\!\{\rmd x\rmd\hat{x}\}~ Q^k[\{x,\hat{x}\}]
M[\{x,\hat{x},\theta,\gamma\}]
&=& C \bra k\ket^k
\label{eq:numerator}
\end{eqnarray}
Integration of (\ref{eq:SPinW1}) over $\{x\}$ reproduces the correct normalization $\int\!\{\rmd x\}W[\{x\}|\{y\}]=1$.
We insert relation (\ref{eq:numerator}) into  (\ref{eq:SPinW1}), which thereby simplifies to
\begin{eqnarray}
W[\{x\}|\{y\}]&=& \sum_{k>0} \frac{p(k)k}{\bra k\ket}
\int\!\{\rmd\hat{x}\} Q^{k-1}[\{x,\hat{x}\}]
\Big\bra M[\{x,\hat{x},\theta\!+\!y,\gamma\}]
\Big\ket_{\!\{\theta,\gamma|k\}}
\end{eqnarray}
Here $\bra \ldots\ket_{\{\theta,\gamma|k\}}$ denotes averaging over the statistics of production and decay rates of those proteins that have $k$ binding partners.
Finally we insert
(\ref{eq:basic_obstacle}) to eliminate the auxiliary kernel $Q[\{x,\hat{x}\}]$, and use $C_k=k!C_0$. This
 brings us after some simple re-arrangements to a transparent
closed equation that involves the physical kernel $W[\{x\}|\{y\}]$ only. This equation, exact in the limit $N\to\infty$, is the final result of the generating functional analysis:
\begin{eqnarray}
W[\{x\}|\{y\}]
&=& \sum_{k\geq 1} \frac{p(k)k}{\bra k\ket} \nonumber
\\[-1mm]
&&\hspace*{-25mm} \times\Big\bra\!\Big\bra
\int\prod_{\ell< k} \left\{\{\rmd x_\ell \rmd y_\ell\}W[\{x_\ell\}|\{y_\ell\}]
 \prod_{\alpha t}\delta\Big[y_{\ell\alpha}(t)\!-\!
\int\!\rmd\tau \sum_{\rho\lambda}
W_{\alpha;\rho\lambda}(\tau|S\bk_\ell)x_{\ell\rho}(t\!-\!\tau)x_{\lambda}(t\!-\!\tau)
\Big]\right\}
\nonumber
\\
&&\hspace*{-25mm}\times
\prod_{\alpha t}\delta\left\{
\rmd x_\alpha(t)
-\dt\Big[\theta_\alpha(t)+y_\alpha(t)-\gamma_\alpha(t) x_\alpha(t)
\bigroom\right.\label{eq:final_result_W}
\\[-1mm]
&&\hspace*{-25mm}\left.~~
+\sum_{\rho\lambda}\int\!\rmd\tau~x_\rho(t\!-\!\tau)\Big(s W_{\alpha;\rho\lambda}(\tau|\bk)
  x_\lambda(t\!-\!\tau)\!+\!\sum_{\ell<k}
W_{\alpha;\rho\lambda}(\tau|\bk_\ell)x_{\ell\lambda}(t\!-\!\tau)\Big)
\Big]
\right\}
\Big\ket_{\!\bk,s;\bk_1\ldots \bk_{k-1}}
\Big\ket_{\!\{\theta,\gamma|k\}}
\nonumber
\end{eqnarray}
From the solution of (\ref{eq:final_result_W}) then follows $D[\{x\}|\{y\}]$, defined in (\ref{eq:meaning2}), via equation
(\ref{eq:Dexpression}), giving
\begin{eqnarray}
D[\{x\}|\{y\}]&=& \sum_{k\geq 0} p(k)
\nonumber
\\
&&\hspace*{-25mm}
\times \Big\bra\!\Big\bra
\int\prod_{\ell\leq k} \left\{\{\rmd x_\ell \rmd y_\ell\}W[\{x_\ell\}|\{y_\ell\}]
 \prod_{\alpha t}\delta\Big[y_{\ell\alpha}(t)\!-\!
\int\!\rmd\tau \sum_{\rho\lambda}
W_{\alpha;\rho\lambda}(\tau|S\bk_\ell)x_{\ell\rho}(t\!-\!\tau)x_{\lambda}(t\!-\!\tau)
\Big]\right\}
\nonumber
\\[-1mm]
&&\hspace*{-25mm}\times
\prod_{\alpha t}\delta\left\{
\rmd x_\alpha(t)
-\dt\Big[\theta_\alpha(t)+y_\alpha(t)-\gamma_\alpha(t) x_\alpha(t)
\bigroom\right.\label{eq:final_result_D}
\\[-1mm]
&&\left.\hspace*{-25mm}~~
+\sum_{\rho\lambda}\int\!\rmd\tau~x_\rho(t\!-\!\tau)\Big(s W_{\alpha;\rho\lambda}(\tau|\bk)
  x_\lambda(t\!-\!\tau)\!+\!\sum_{\ell\leq k}
W_{\alpha;\rho\lambda}(\tau|\bk_\ell)x_{\ell\lambda}(t\!-\!\tau)\Big)
\Big]
\right\}
\Big\ket_{\!\bk,s;\bk_1\ldots \bk_k}
\Big\ket_{\!\{\theta,\gamma|k\}}
\nonumber
\end{eqnarray}
Although different at the level of mathematical details, the structure of the above equations is very similar to what was found in  non-equilibrium statistical mechanical studies of spin models on finitely connected random graphs, such as \cite{Hatchett_etal,MozeikaCoolen,MimuraCoolen1,MimuraCoolen2}.
In  the special case of protein interaction networks with Poissonian degree distributions, i.e. $p(k)=\rme^{-c}c^k/k!$, where $p(k\!+\!1)(k\!+\!1)/\bra k\ket=p(k)$ for all $k$, a simple transformation $k\to k\!+\!1$ shows that the kernels $P[\{x\}|\{y\}]$ and  $D[\{x\}|\{y\}]$ will be identical.
In all other cases this will not be true.

\subsection{Cavity interpretation of the order parameter equations}

\begin{figure}[t]
\thicklines
\unitlength=0.14mm
\hspace*{25mm}
\begin{picture}(350,280)
\put(140,250){[A]}
\put(100,150){\blacknode}\put(100,110){ $i$} \put(100,180){$\{x_i\}$}
\put(200,150){\node}
\put(114,151){\vector(1,0){71}} \put(185,151){\vector(-1,0){71}}
\put(86,151){\vector(-1,0){90}} \put(-4,151){\vector(1,0){90}}
\put(91,160){\vector(-1,1){60}} \put(31,220){\vector(1,-1){60}}
\put(91,140){\vector(-1,-1){60}} \put(31,80){\vector(1,1){60}}
\put(210,160){\vector(1,1){60}} \put(270,220){\vector(-1,-1){60}}
\put(210,140){\vector(1,-1){60}}\put(270,80){\vector(-1,1){60}}
\put(-18,150){\node} \put(20,230){\node} \put(20,70){\node}
\put(280,70){\node} \put(280,230){\node}
\end{picture}
\hspace*{10mm}
\begin{picture}(350,280)
\put(140,250){[B]}
\put(100,150){\blacknode}\put(100,110){$i$} \put(100,180){$\{x_i\}$}
\put(200,150){\node}\put(200,150){\smallnode}
\put(195,110){$j$}
\put(235,145){$\theta_j^\alpha\!+y_j^\alpha(\{x_i,x_j\})$}
\put(186,151){\vector(-1,0){72}}
\put(-4,151){\vector(1,0){90}}
\put(31,220){\vector(1,-1){60}}
\put(31,80){\vector(1,1){60}}
\put(210,160){\vector(1,1){60}}
\put(270,220){\vector(-1,-1){60}}
\put(210,140){\vector(1,-1){60}}\put(270,80){\vector(-1,1){60}}
\put(-18,150){\node} \put(-18,150){\smallnode}
\put(21,229){\node} \put(21,229){\smallnode}
\put(21,71){\node} \put(21,71){\smallnode}
\put(280,70){\node} \put(280,230){\node}
\end{picture}
\\
\hspace*{55mm}
\begin{picture}(350,270)
\put(140,250){[C]}
\put(100,150){\blacknode}\put(100,110){$i$} \put(100,180){$\{x_i\}$}
\put(200,150){\node}\put(200,150){\smallnode}
\put(195,110){$j$}\put(235,145){$\theta_j^\alpha\!+y_j^\alpha(\{x_i,x_j\})$}

\put(240,230){$m$}\put(310,230){$\theta_m^\alpha\!+y_m^\alpha(\{x_j,x_m\})$}
\put(242,60){$n$}\put(310,60){$\theta_n^\alpha\!+y_n^\alpha(\{x_j,x_n\})$}

\put(-4,151){\vector(1,0){90}}
\put(31,220){\vector(1,-1){60}}
\put(31,80){\vector(1,1){60}}
\put(270,220){\vector(-1,-1){60}}
\put(270,80){\vector(-1,1){60}}
\put(-18,150){\node} \put(-18,150){\smallnode}
\put(21,229){\node} \put(21,229){\smallnode}
\put(21,71){\node} \put(21,71){\smallnode}
\put(280,70){\node} \put(280,230){\node}
\end{picture}

\vspace*{-5mm}
    \caption{
      Interpretation of our order parameter equations. Top left [A]: section of the original interaction network,  showing a site $i$ which has four interaction partners (these define the set $\partial_i$). Top right [B]: a locally modified network where information flow from $i\to\partial_i$ is prohibited (so node $i$ is effectively removed), but the production rates of {\em all} $j\in\partial_i$ (i.e. of all sites marked as ${\bigcirc\hspace*{-2.8mm}\circ}$~) are adjusted in compensation, so that none of the concentrations in the system change, following  (\ref{eq:production_change}). Bottom [C]: result of a further modification, where also the information flow from $j\to \partial_j$ is prohibited (so also node $j$ is effectively removed), but again compensated for by appropriate adjustment of the production rates of the sites in $\partial_j=\{m,n\}$.
    }
    \label{fig:interpretation}
\end{figure}
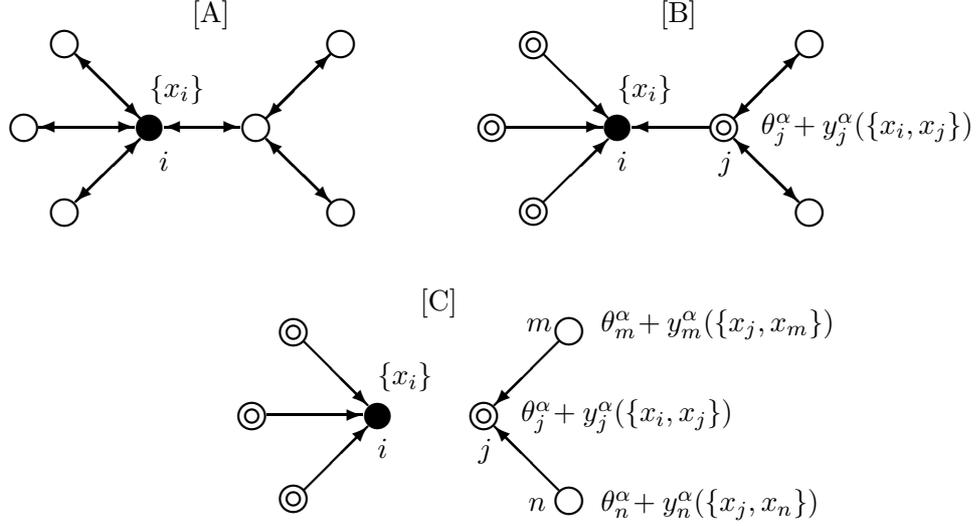

One can understand our results (\ref{eq:final_result_W},\ref{eq:final_result_D}) at a more intuitive level, starting from equations
(\ref{eq:microdynamics}). Any protein species $i$ interacts with the proteome only via its direct partners, the species in the set $\partial_i=\{\ell|~c_{i\ell}=1\}$. Given the paths $\{x_\ell\}$ taken by the concentrations of the species $\ell\in\partial_i$, the path $\{x_i\}$ taken by $i$ could in principle be calculated (modulo non-invertibility issues) by solving a linear equation, giving an expression of the form $\{x_i\}=\F_{i}[\{x_\ell\},\ell\!\in\!\partial_i]$. This is illustrated in figure \ref{fig:interpretation}[A], where site $i$ is shown in black for an example with $k_i=4$. Next we imagine changing the structure of the graph locally, by removing the information flow from $i$ to $\partial_i$, see figure \ref{fig:interpretation}[B] where the protein species in $\partial_i$ are drawn as ${\bigcirc\hspace*{-2.7mm}\circ}$~. We compensate for this intervention, however, by adjusting the production rates of all $j\in\partial_i$ according to $\theta_j^\alpha(t)\to \theta_j^\alpha(t)+y_j^\alpha(\{x_i,x_j\},t)$ such that we do not change any of the concentration paths in the system, which requires
\begin{eqnarray}
y_j^\alpha(\{x_i,x_j\},t)&=& \int\!\rmd s\sum_{\rho\lambda}W_{\alpha;\rho\lambda}(t-s|\bk_{i\ell})x_i^\lambda(s)x_j^\rho(s)
\label{eq:production_change}
\end{eqnarray}
 Hence, with the conventions $\partial_i=\{j_{i1},\ldots,j_{ik_i}\}$ and $D[\{x\}]=D[\{x\}|\{0\}]$ we may write
\begin{eqnarray}
D[\{x\}]&=& \lim_{N\to\infty}\frac{1}{N}\sum_i\overline{\delta\big[\{x\}-\F_{i}[\{x_{j_{i1}},\ldots,x_{j_{ik_i}}\}]\big]}
\nonumber
\\
&=&\lim_{N\to\infty}\frac{1}{N}\sum_i\overline{\int\!\prod_{\ell=1}^{k_i}\Big\{\{\rmd x_\ell\} \delta[\{x_\ell\}\!-\!\{x_{j_{i\ell}}\}]
\Big\}~\delta\big[\{x\}\!-\!\F_{i}[\{x_1,\ldots,x_{k_i}\}]\big]}
\nonumber
\\
&&\hspace*{-18mm} =\lim_{N\to\infty}\frac{1}{N}\sum_i\overline{\int\!\prod_{\ell=1}^{k_i}\Big[\{\rmd x_\ell\}
\delta[\{x_\ell\}\!-\!\{x_{j_{i\ell}}\}]_{i~{\rm removed},~\{\theta_{j_{i\ell}}+y_{j_{i\ell}}(\{x,x_\ell\})\}}
\Big]~\delta\big[\{x\}\!-\!\F_{i}[\{x_1,\ldots,x_{k_i}\}]\big]}
\nonumber
\\
&& \hspace*{-18mm} =\lim_{N\to\infty}\frac{1}{N}\sum_i\int\!\prod_{\ell=1}^{k_i}\Big[\{\rmd x_\ell \rmd y_\ell\}\Big]
\\
&&\hspace*{-10mm}
\times\overline{
\prod_{\ell=1}^{k_i}\Big[\delta[\{x_\ell\}\!-\!\{x_{j_{i\ell}}\}]_{i~{\rm removed},~\{\theta_{j_{i\ell}}+y_\ell\}}
\delta[\{y_\ell\}-\{y_{j_{i\ell}}(\{x,x_\ell\})\}]
\Big]\delta\big[\{x\}\!-\!\F_{i}[\{x_1,\ldots,x_{k_i}\}]\big]}
\nonumber
\end{eqnarray}
If the original network had no loops, then in the new `cavity' graph the branches attached to sites $\ell\in\partial_i$ are disconnected, and each will behave as an independent graph, for large $N$ topologically equivalent to the original. It follows that in the cavity graph
we can treat the paths $\{x_\ell\}$ with $\ell\in\partial_i$ as statistically independent, and assume they have identical statistical properties, so
\begin{eqnarray}
D[\{x\}]
&=&\lim_{N\to\infty}\frac{1}{N}\sum_i\int\!\prod_{\ell=1}^{k_i}\Big[\{\rmd x_\ell \rmd y_\ell\}\overline{\Big(\frac{1}{N}\sum_j\delta[\{x_\ell\}\!-\!\{x_{j}\}]_{i~{\rm removed},~\{\theta_{j}+y_\ell\}}\Big)}\Big]
\\
&&
\times\overline{
\delta\big[\{x\}\!-\!\F_{i}[\{x_1,\ldots,x_{k_i}\}]\big]\prod_{\ell\leq k_i}\delta[\{y_\ell\}-\{y_{j_{i\ell}}(\{x,x_\ell\})\}]
}
\nonumber\\
&=&\lim_{N\to\infty}\frac{1}{N}\sum_i\int\!\prod_{\ell=1}^{k_i}\Big[\{\rmd x_\ell \rmd y_\ell\}W[\{x_\ell\}|\{y_\ell\}]\Big]
\\
&&
\times~\overline{
\delta\big[\{x\}\!-\!\F_{i}[\{x_1,\ldots,x_{k_i}\}]\big]\prod_{\ell\leq k_i}\delta[\{y_\ell\}-\{y_{j_{i\ell}}(\{x,x_\ell\})\}]
}
\nonumber
\end{eqnarray}
One will now be led to equation (\ref{eq:final_result_D}), upon simply adding $\{y\}$ to the production rates $\{\theta_i\}$ (which brings us from $D[\{x\}]$ to $D[\{x\}|\{y\}]$) and  with the last two lines of  (\ref{eq:final_result_D})
representing the more explicit representation of our symbolic expression $\delta[\{x\}\!-\!\F_{i}[\{x_1,\ldots,x_{k_i}\}]]$.

What remains is to understand the origin of (\ref{eq:final_result_W}), which requires an expression for the concentration path statistics of nodes in $\partial_i$, such as $j$ in figure \ref{fig:interpretation}[B]. We repeat the process of blocking the information flow away from the node of which we try to calculate the concentration paths, while compensating the production rates of its partners appropriately. Now this means removing $j$, and adjusting the production rates of the nodes in $\partial_j$ (except for $i$, which has been removed already); see figure \ref{fig:interpretation}[C].
The equation for $W[\{x\}|\{y\}]$ is therefore nearly identical to that of $D[\{x\}|\{y\}]$, with two differences:
first, since we are looking strictly at nodes $j$ that were connected to a (now removed) cavity node $i$, these nodes $j$ no longer have in-degree statistics $p(k)$, and second, only $k_j-1$ of the original $k_j$ partners of each $j$ contribute to $\{x_j\}$. Since our random graph ensemble has no degree-degree correlations, the modified degree statistics $\tilde{p}(k)$ of the nodes $j\in\partial_i$ that were initially attached to (randomly drawn) `cavity' sites $i$ follows from
\begin{eqnarray}
\tilde{p}(k)=\lim_{N\to\infty}\frac{\sum_{ij}c_{ij}\delta_{k_j,k}}{\sum_{ij}c_{ij}}=
\lim_{N\to\infty}\frac{k}{N\bra k\ket}\sum_j\delta_{k_j,k} =\frac{p(k)k}{\bra k\ket}
\end{eqnarray}
 We can thus obtain our equation for $W[\{x\}|\{y\}]$ by making in the right-hand side of (\ref{eq:final_result_D}) the replacements $p(k)\to p(k)k/\bra k\ket$, $\prod_{\ell\leq k}\to \prod_{\ell<k}$, and $\sum_{\ell\leq k}\to\sum_{\ell<k}$. The result is indeed (\ref{eq:final_result_W}).

It is satisfactory that we have been able to use generating functional analysis (GFA) to obtain (\ref{eq:final_result_W},\ref{eq:final_result_D}), which is  more precise and direct than the above reasoning and did not require the strict absence of loops
(in effect, GFA confirms that for $N\to\infty$ any loops generated in (\ref{eq:ensemble}) have vanishing impact on the process). Second, as soon as we use more complicated ensembles than (\ref{eq:ensemble}), e.g. those with degree-degree correlations as in \cite{CoolenPerez2008}, or introduce correlations between the reaction rates of distinct protein pairs, the above simple arguments would become prohibitively messy, whereas the GFA route should in principle remain open.

\section{Solution of the macroscopic equations}

The focus must now turn to solving equation (\ref{eq:final_result_W}) for $W[\{x\}|\{y\}]$, from which $D[\{x\}|\{y\}]$ follows via
(\ref{eq:final_result_D}). This is a highly nontrivial problem, on which progress has so far been slow; not just here, but in all GFA studies of processes on finitely connected graphs \cite{Hatchett_etal,MimuraCoolen1,MimuraCoolen2}. The possible routes to be explored come, roughly, in four areas. This paper is only the first step in a research programme, establishing `proof of principle' that GFA methods can indeed be used to study the proteome, hence in view of space limitations we will here only comment briefly on each area:

\subsection{Numerical solution of the macroscopic laws}
Due to the exponential increase with time of the number of macroscopic observables concerned, even in studies with discrete time and discrete variables \cite{Hatchett_etal,MimuraCoolen1,MimuraCoolen2}, numerical solution of equations such as (\ref{eq:final_result_W}) was possible only for a small number of time steps. In the present problem, where the arguments of $W[\{x\}|\{y\}]$ are continuous paths, numerical solution is not a realistic option.

\subsection{Working out solutions in specific simplifying limits}
The two main limits where analytical solution is possible are low connectivity and high connectivity (where $\bra k\ket\to \infty$ must be preceded by a suitable re-scaling $k_{\alpha\beta}^+\to k_{\alpha\beta}^+/\bra k\ket$ of all on-rates). If we assume our network has no disconnected nodes, then the lowest overall connectivity is found for $p(k)=\delta_{k1}$,
where we find (\ref{eq:final_result_W},\ref{eq:final_result_D}) reducing to
\begin{eqnarray}
W[\{x\}|\{y\}]
&\!=\!& \Big\bra\!\Big\bra
\prod_{\alpha t}\delta\Big\{
\rmd x_\alpha(t)
-\dt\Big[\theta_\alpha(t)+y_\alpha(t)-\gamma_\alpha(t) x_\alpha(t)
\\
&&\hspace*{-0mm}~~
+\sum_{\rho\lambda}\int\!\rmd\tau~x_\rho(t\!-\!\tau)\Big(s W_{\alpha;\rho\lambda}(\tau|\bk)
  x_\lambda(t\!-\!\tau)\Big)
\Big]
\Big\}
\Big\ket_{\!\bk,s}
\Big\ket_{\!\{\theta,\gamma\}}
\nonumber
\\
D[\{x\}|\{y\}]&\!=\!& \Big\bra\!
\int\!\{\rmd x^\prime \rmd y^\prime\}W[\{x^\prime\}|\{y^\prime\}]
 \prod_{\alpha t}\delta\Big[y^\prime_{\alpha}(t)\!-\!
\int\!\rmd\tau \sum_{\rho\lambda}
W_{\alpha;\rho\lambda}(\tau|S\bk^\prime)x^\prime_{\rho}(t\!-\!\tau)x_{\lambda}(t\!-\!\tau)
\Big]
\nonumber
\\
&&\hspace*{-0mm}\times\Big\bra
\prod_{\alpha t}\delta\Big\{
\rmd x_\alpha(t)
-\dt\Big[\theta_\alpha(t)+y_\alpha(t)-\gamma_\alpha(t) x_\alpha(t)
\\
&&
\hspace*{-10mm}~~
+\sum_{\rho\lambda}\int\!\rmd\tau~x_\rho(t\!-\!\tau)\Big(s W_{\alpha;\rho\lambda}(\tau|\bk)
  x_\lambda(t\!-\!\tau)\!+\!
W_{\alpha;\rho\lambda}(\tau|\bk_\ell)x^\prime_{\lambda}(t\!-\!\tau)\Big)
\Big]
\Big\}
\Big\ket_{\!\bk,s;\bk^\prime}
\Big\ket_{\!\{\theta,\gamma\}}
\nonumber
\end{eqnarray}
These equations are easily interpreted (all cavity sites are now isolated) and
 converted into a pair of coupled stochastic differential equations, whose
statistics represent the diversity of concentration paths in the original $N$-protein system.
For large connectivity we first re-scale the reaction on-rates according to $k^{\alpha\beta^+}=
\tilde{k}^{\alpha\beta +}/\bra k\ket$. One now finds that $D[\{x\}]=D[\{x\}|\{0\}]=W[\{x\}|\{0\}]$
obeys a closed equation as $\bra k\ket\to \infty$, which using the
law of large numbers becomes
\begin{eqnarray}
D [\{x\}]&=&\Big\langle\prod_{\alpha t}\delta\Big[\rmd x_{\alpha}(t)-\rmd t\bigg(\theta_\alpha(t)-\gamma_\alpha(t)x_\alpha(t)+s\sum_{\rho\lambda}\int\! \rmd\tau~ W_{\rho\lambda}(\tau|\textbf{k})x_{\rho}(t\!-\!
\tau)x_\lambda(t\!-\!\tau)
\nonumber\\
&& \hspace{8mm}+\sum_{\rho\lambda}\int\! \rmd\tau~ W_{\rho\lambda}(\tau|S\tilde{\bk})x_\lambda(t\!-\!\tau)\int\!\{\rmd x^\prime\}~ D[\{x^\prime\}]x^\prime_{\rho}(t\!-\!\tau)\bigg)\Big]\Big\rangle_{s,\bk,\tilde{\bk},\{\theta, \gamma\}}
\label{eq:meanfield}
\end{eqnarray}
Again the problem can be converted into a relatively simple stochastic equation.
The protein-protein interaction occurs in a `mean field' way. In fact for $\bra k\ket\to\infty$ we can extract from (\ref{eq:meanfield}) a closed equation for the disorder-averaged path $\{X\}=\int\!\{\rmd x\}D[\{x\}]\{x\}$, rather than a closed theory in the language of correlation- and response functions (see e.g. \cite{CrisantiSompolinsky1987}) which one would have found upon taking the diverging connectivity limit in models of disordered spin systems or neural networks on finitely connected random graphs.

\subsection{Constructing ad-hoc approximations}

Various approximations could be considered.
First, one could solve (\ref{eq:final_result_W}) iteratively, generating a sequence of measures $W_n[\{x\}|\{y\}]$ by substitution for each $n$ of
$W_n[\{x\}|\{y\}]$ in the right-hand side of (\ref{eq:final_result_W}) and defining the left-hand side as $W_{n+1}[\{x\}|\{y\}]$. Upon starting e.g. from the trivial $W_0[\{x\}|\{y\}]=\delta[\{x\}]$, one would find $W_1[\{x\}|\{y\}]$ describing a dimers-only system of proteins, and $W_{2}[\{x\}|\{y\}]$ describing interacting proteins but without an Onsager reaction term, etc.
An alternative would be a linear response approximation, based on truncating  the exact expansion
\begin{eqnarray}
W[\{x\}|\{y\}]=W[\{x\}]+\sum_\alpha\!\int\!\rmd t~ y_\alpha(t)\frac{\delta W[\{x\}]}{\delta\theta_\alpha(t)}+
\frac{1}{2}\sum_{\alpha\beta}\!\int\!\rmd t\rmd t^\prime~ y_\alpha(t)y_\beta(t^\prime)\frac{\delta^2 W[\{x\}]}{\delta\theta_\alpha(t)\delta\theta_\beta(t^\prime)}+\ldots
\nonumber
\\[-4mm]
\label{eq:expandW}
\end{eqnarray}
Each such approximation will have specific strengths and weaknesses, which will depend on the statistical characteristics (topology, strength of  interactions) of the system being studied.

\subsection{Probing further the mathematical structure of the macroscopic laws}

An obvious question to be investigated is whether in stationary states  there exist exact closed laws for a reduced set of static order parameters
(this may require the equivalent of detailed balance, e.g. demanding that $S\bk=\bk$ for all reaction rates). The natural candidates would be the probability densities $W(\bx|\by)$ for asymptotic time-averages $\bx$ of protein concentrations, given asymptotic time averages $\by$ of production perturbations (as opposed to paths).

 One could also try to exploit the origin of (\ref{eq:final_result_W}) as a saddle-point equation for a functional.
 Upon applying the various transformations of order parameters directly to the function $\Psi[\{P,Q\}]+\Phi[\{P\}]+\Omega[\{Q\}]$ in (\ref{eq:saddlepointform}), with an inverse Fourier transform for $Q$, one obtains
 \begin{eqnarray}
{\cal L}[\{V,W\}]
&=&\frac{1}{2}-\int\!\{\rmd x \rmd y\}~W[\{x\}|\{y\}]
V[\{y\}|\{x\}]
\nonumber
\\
&&
+\frac{1}{2}\!\int\!\{\rmd x\rmd y\rmd x^\prime\! \rmd y^\prime\}~
W[\{x\}|\{y\}]~ {\cal M}[\{x,y\};\!\{x^\prime,y^\prime\}]~ W[\{x^\prime\}|\{y^\prime\}]
\label{eq:L_VW}
\\
&&\hspace*{-10mm} +\sum_k \frac{p(k)}{\bra k\ket}
\Big\bra\log
\int\!\{\rmd x\}\!\! \int\!\prod_{\ell\leq k}
\Big[\{\rmd y_\ell\}~V[\{y_\ell\}|\{x\}]\Big]
\Big\bra\delta\Big[\{x\}\!-\!F\big[\{\theta\!+\!\sum_{\ell\leq k} \!y_\ell\};s,\bk\big]\Big]
\Big\ket_{\!\bk,s}
\Big\ket_{\!\{\theta,\gamma|k\}}
\nonumber
\end{eqnarray}
with a kernel ${\cal M}[\ldots;\ldots]$ that contains all the information on protein-protein interaction,
\begin{eqnarray}
{\cal M}[\{x,y\};\!\{x^\prime,y^\prime\}]&=& \Big\bra
\prod_{\alpha t}\delta\Big[y_\alpha(t)\!-\!
\int\!\rmd\tau\! \sum_{\rho\lambda} W_{\alpha;\rho\lambda}(\tau|\bk)x_\rho(t\!-\!\tau)x^\prime_\lambda(t\!-\!\tau)
\Big]
\nonumber
\\
&&\times
\prod_{\alpha t}\delta\Big[y^\prime_\alpha(t)\!-\!
\int\!\rmd\tau\! \sum_{\rho\lambda} W_{\alpha;\rho\lambda}(\tau|S\bk)x^\prime_\rho(t\!-\!\tau)x_\lambda(t\!-\!\tau)\Big]
\Big\ket_{\bk}
\label{eq:kernel_M}
\end{eqnarray}
and where $F[\{\theta\};s,\bk]$ denotes the solution of the equation
\begin{eqnarray}
\frac{\rmd}{\rmd t}
x_\alpha(t)&=& \theta_\alpha(t)-\gamma_\alpha(t) x_\alpha(t)+ s\sum_{\rho\lambda}\int\!\rmd\tau~W_{\alpha;\rho\lambda}(\tau|\bk)
 x_\rho(t\!-\!\tau)x_\lambda(t\!-\!\tau)
 \end{eqnarray}
One's instinct next would be to attempt a  variational formulation: to define on the basis of (\ref{eq:L_VW}) a function ${\cal L}[\{W\}]$ whose minimum would give the true solution. It turns out that this will never be possible here, due to an intriguing property of (\ref{eq:L_VW}): one can show that  for {\em all} functions that obey causality, the surface ${\cal L}[\{V,W\}]$ is no longer dependent upon whether or not the proteins interact.
The conclusion must be that all attempts to convert (\ref{eq:L_VW}) into a variational problem, suitable for generating variational approximations of $W[\{x\}|\{y\}]$, are doomed from the start, since variation within any subset of causal measures will at most give us information on $W[\{x\}|\{y\}]$ that would be true irrespective of whether or not proteins form hetero-dimers.

\vspace*{\fill}

\section*{Ackowledgements}

ACCC would like to thank the Engineering and Physical Sciences Research Council (UK) for support
in the form of a Springboard Fellowship.

\end{document}